\documentclass[12pt]{article}
\usepackage[linktocpage=true,plainpages=false,pdfpagelabels=false]{hyperref}
\usepackage{geometry}
\usepackage[usenames]{color}
\usepackage{color}
\usepackage{doi} 
\usepackage{amsmath,amssymb}
\usepackage{cite}
\usepackage{array}
\usepackage{multicol}
\usepackage{graphicx}
\usepackage{lmodern}
\usepackage{textcomp}
\usepackage[T1]{fontenc}
\definecolor{linkcolor}{rgb}{0.6,0,0}
\definecolor{citecolor}{rgb}{0,0.5,0}
\definecolor{urlcolor}{rgb}{0,0,1}
\hypersetup{colorlinks, linkcolor={linkcolor},
    citecolor={citecolor}, urlcolor={urlcolor}}
\begin{document}
\title{Friedmann cosmology in Regge-Teitelboim gravity}
\author{A. A. Sheykin\thanks{E-mail: a.sheykin@spbu.ru},
S. A. Paston\thanks{E-mail: paston@pobox.spbu.ru}\\
{\it Saint Petersburg State University, St.-Petersburg, Russia}
}
\date{\vskip 15mm}
\maketitle

\begin{abstract}
This paper is devoted to the approach to gravity as a theory of a surface embedded in a flat ambient space. After the brief review of the properties of original theory by Regge and Teitelboim we concentrate on its field-theoretic reformulation, which we call splitting theory. In this theory embedded surfaces are defined through the constant value surfaces of some set of scalar fields in high-dimensional Minkowski space. We obtain an exact expressions for this scalar fields in the case of Friedmann universe. We also discuss the features of quantisation procedure for this field theory.

\end{abstract}
\newpage
\section{Introduction}	
The question about the possible modifications of general relativity (GR) arose soon after its appearance. First modifications of GR (Weyl, Eddington, Einstein itself etc.) were conducted for the purpose of unifying all then known field theories, such as gravitational and electromagnetic. Among these attempts we should mention multidimensional theories (Kaluza, Klein, Mandel, Fock), in which an additional fifth dimension was introduced to describe the electromagnetism. A detailed historical study of these theories can be found in the monograph by V. P. Vizgin.\cite{vizgin} The popularity of these theories has faded in the early 50s, when two additional interactions were discovered: they did not fit into this scheme. In addition, the quantum theory, which has been developed at the same time, was much more fruitful and attractive for scientists. In the context of multidimensionality the studies of Yu. B. Rumer\cite{rumer} must though be noted, in which the author attempted to interpret the quantum theory as a fundamentally five-dimensional theory.

However, many new issues were found during The Golden Age of GR. First of all, it became apparent that the quantization of GR in its standard metric form is fraught with insurmountable difficulties.\cite{carlip} This resulted in the search of alternative formulations of the gravity theory which would be more suitable for quantization.
Furthermore, the first discrepancies between experimental observations and the predictions of GR were discovered. These discrepancies include, in particular, the problem of hidden mass in the Universe, which resulted in the Dark Matter hypothesis, and the cosmological constant problem, which resulted in the proposition of the Dark Energy existence. These two main problems led to the search of modifications of GR which would explain them. Besides that, it was hoped that the modification of gravity could solve certain problems of the particle physics, e.g. the mass hierarchy problem.

It is surprisingly enough that these completely different research paths at a certain step resulted in one idea: the consideration of our 4-dimensional spacetime as a surface in a spacetime of higher dimension. In contrast to Kaluza-Klein scheme, extra dimensions in this approach are not necessarily small.

The approach to gravity as a dynamics of a surface embedded in some curved spacetime is usually called the braneworld and includes such models as ADD-brane, DGP-brane, RS-brane, Rubakov-Shaposhnikov thick brane etc.\cite{barvin} This approach was initially considered as phenomenology, but it was shown later that many braneworld models emerge as low-energy limits of string theory. In the framework of braneworld the problems of particle physics and non-GR gravity phenomena are mainly treated.

The search for new formulation of gravity which would be more suitable for quantization, in its turn, leads to the creation of the so called embedding theory --- Regge-Teitelboim (RT) gravity, in which gravity is treated as a dynamics of a surface isometrically embedded in a flat ambient spacetime of higher dimension. The authors of this approach were inspired by the successful use of isometric embedding in other theories, e.g. relativistic mechanics ($1D$ surface in $4D$ spacetime), string theory ($2D$ spacetime in $D$-dim Minkowski spacetime), ADM gravity etc. One of the most serious problems with quantization of gravity in its standard form is the absence of a flat spacetime to which we are accustomed to deal in the quantization process. The natural appearance of a flat spacetime in RT approach could simplify the canonical formulation of gravity and help to avoid some ideological problems of quantum gravity, such as the definition of causality, the preferred time direction and so on. It should be noted though that RT approach is not the only possible theory of gravity based on the isometrical embeddings: there are some theories whose dynamics is driven by Gauss-Codazzi-Ricci equations.\cite{maia,maia1,maia7} It is also possible to include the ideas of Regge and Teitelboim in the framework of braneworld. The Friedmann-like cosmological models were discussed\cite{davgur06} in the context of such an unified theory.
Friedmann cosmology is also the main subject of interest of the present work. 

	\section{Friedmann cosmology in Regge-Teitelboim approach}

As in RT approach the spacetime is treated as $4D$ surface isometrically embedded in a flat ambient space, the embedding function $y^a$ of this surface becomes a dynamical variable instead of the metric $g_{\mu\nu}$:
\begin{align}
	g_{\mu\nu}(x) = \partial_\mu y^a(x) \partial_\nu y^b(x) \eta_{ab},\label{metric}
\end{align}
where $\mu,\nu = 0 \ldots 3$, $\eta_{ab}$ ia an $N$-dimensional Minkowski metric, $a,b = 0 \ldots N-1$. For a general $4D$ metric $N=10$. If the symmetry of the metric is high enough, $N$ can be smaller. For open and closed and spatially flat FRW metric:
\begin{align}
		ds^2_c=dt^2-a^2(t)(d\chi^2+\sin\text{h}^2\chi(d\theta^2+\sin^2\theta d\phi^2))
\end{align}
$N=5$ and the embedding function looks like
 \begin{align}
	\begin{aligned}[b]
		&y^0= \int dt \sqrt{\dot{a}^2+1},\\
		&y^1= a(t)\, \cos \chi,  \\
		&y^2= a(t)\, \sin \chi \cos \theta, \\
		&y^3= a(t)\, \sin \chi \sin \theta \cos \phi, \\
		&y^4= a(t)\, \sin \chi \sin \theta \sin \phi,
	\end{aligned}
	\qquad
	\begin{aligned}[b]
		&y^0= a(t)\, \cosh \chi,   \\
		&y^1= a(t)\, \sinh \chi \cos \theta, \\ \label{2.11}
		&y^2= a(t)\, \sinh \chi \sin \theta \cos \phi, \\
		&y^3= a(t)\, \sinh \chi \sin \theta \sin \phi,\\
		&y^4= \int dt \sqrt{\dot{a}^2-1}, 
	\end{aligned}
\end{align}
for open and closed models respectively. The embedding for a spatially flat model:
\begin{align}
		ds^2=dt^2-a^2(t)(dr^2+r^2(d\theta^2+\sin^2\theta d\phi^2))
\end{align}
 is a bit more complicated:
\begin{align}
				&y^{0,1}= \dfrac{1}{2}\left(a(t) r^2 + \int \dfrac{dt}{\dot{a}} \pm a(t) \right)\!, \ y^2= a(t)\, r \cos \theta, \label{flat}\nonumber \\
				&y^3= a(t)\, r \sin \theta \cos \phi, \ y^4= a(t)\, r \sin \theta \sin \phi.
\end{align}
Here and hereafter $a(t)$ is a scale factor, the ambient space signature is $(+----)$.

These embeddings were found in 1933.\cite{robertson1933} It can be shown\cite{statja29} that they are the only embeddings with  $N=5$ which inherit the symmetry of the FRW metric ($SO(4)$, $SO(1,3)$ and $SO(3)\triangleright T^3$ respectively). Despite the fact that the $t={const}$ surfaces of a spatially flat embedding are not 3D planes, they are $SO(3)\triangleright T^3$-symmetric. The properties of this embedding are similar to those of the Fujitani-Ikeda-Matsumoto embedding of the Schwarzchild metric.\cite{statja27} 

The dynamics of the embedding theory is driven by RT equations:
\begin{align}
&G^{\mu\nu}-\varkappa T^{\mu\nu}=	\varkappa \tau^{\mu\nu}, \label{2.5} \\
&\partial_{\mu}(\sqrt{-g} \tau^{\mu\nu} \partial_{\nu} y^a)  = 0, \label{2.6}
\end{align}	
where \eqref{metric} is substituted in the Einstein tensor $G^{\mu\nu}$ and EMT $T^{\mu\nu}$ of matter on the surface. The tensor $\tau_{\mu\nu}$ can be interpreted as an additional term in EMT which leads to a deviation of predictions of the theory from GR. Since $\tau^{\mu\nu}$ extends the Einsteinian dynamics, one can consider it as a possible dark matter candidate. Concurrently, the nonlinearizability\cite{deser} of RT equations complicates the quantization significantly, and the <<extra solutions>> thus can be treated as an artefact of a theory which must be eliminated.

The RT equations are more difficult to solve than the Einstein ones because of their highly nonlinear structure. However, in the presence of high symmetry (e.g. the Friedmann one) they are reduced to ODE and can be solved.

The calculation performed within the framework of Friedmann symmetry showed\cite{statja26} that in the assumption of inflationary expansion in the early Universe the <<extra solutions>> turn out to be exponentially suppressed after inflation and do not contribute significantly to the predictions of the theory. The RT gravity at the large scales is thus equivalent to GR that potentially simplifies the quantization.

\section{Field-theoretic reformulation}
Besides the fact that in RT approach some quantization issues can be avoided, it nevertheless inherits certain disadvantages from GR; e.g. the nesessity of introduction of the coordinates $x^\mu$ on the surface. Because of that a field-theoretic reformulation of RT gravity was proposed\cite{statja25} (we call it splitting theory), the main idea of which is the following.

Let us define a set of scalar fields $z^A$, $A=1\ldots N-4$ in an $N$-dimensional Minkowsky spacetime. The surfaces of constant value of these fields $z^A(y^a)=\text{const}$ define a family of $4D$ surfaces without the need of any coordinate system on them. In such a manner all invariant geometrical characteristics of the surfaces can be written through $z^A(y^a)$ instead of $y^a(x^{\mu})$.

In particular, one can define surfaces with given metric through fields $z^A$. According to \eqref{2.11} and \eqref{flat}, in the case of FRW-metric it is sufficient to use the field $z$ with only one component. In particular, for closed model in the inflational and radiation-dominated epochs $z$ has the form
\begin{align}
z(y^a)=\begin{cases}
	&y^0-\sqrt{a^2-\frac{3}{\varkappa\rho_{infl}}},\qquad \sqrt{\frac{3}{\varkappa\rho_{infl}}}\le a \le a_{i-r},\\
	&y^0+\sqrt{\frac{\varkappa k_{rad}}{3}}\arccos\left(a\sqrt{\frac{3}{\varkappa k_{rad}}}\right)+C,
	\quad a \ge a_{i-r},
\end{cases}
\end{align}
where $a=\sqrt{{y^1}^2+{y^2}^2+{y^3}^2+{y^4}^2}$;  $k_{rad}=\rho_{rad}a^4=const$, $\rho_{infl}=const$ and $\rho_{rad}$ is the energy density of inflaton and radiation; $a_{i-r}$ is the value of $a$ at the end of inflation epoch and $C$ is a constant which provides a continuity
of $z(y^a)$ at $a=a_{i-r}$.

An important feature of the above theory is the existence of certain invariance analogous to the diffeomorphisms in GR:

\begin{align}
	z^A(y)\rightarrow z'^A(y)=f^A(z(y)), \label{5.6}
\end{align}
which can be interpreted as the invariance under renumbering of surfaces (defining the surfaces with the same geometrical characteristics through different configurations of $z^A$). The solutions of dynamical equations of the theory must therefore respect this invariance, which put certain restrictions on the action of the theory.\cite{nuclphys}

However, it must be noted that the action of the theory can be chosen in the form analogous to the Einstein-Hilbert one:
\begin{align}
	S=-\frac{1}{2\varkappa}\int d^{N}y \sqrt{|w|} (R-2\varkappa \mathcal L_m), 	\label{act}
\end{align}
where $y^a$ are lorenzian Minkowski space coordinates, $\mathcal L_{m}$ is a lagrangian of matter on the surface, $R$ is a Ricci scalar of the constant value surfaces of $z^A$, $w=\det w^{AB}$, $w^{AB}=\partial_a z^A \partial_b z^{B} \eta^{ab}$ is a quantity which plays the role of metric in the directions orthogonal to the surfaces. In this case the field equations for each surface are equivalent to the RT ones:
\begin{align}
	b^A\,_{ab} (G^{ab} -\varkappa T^{ab} )=0,\label{RT_split}
\end{align} 
where $G^{ab}=G^{\mu\nu}\partial_\mu y^a\partial_\nu y^b$ (same for $T^{ab}$), $b^A\,_{ab}$ is a second fundamental form of the surface. The above quantities are written in a coordinate-invariant way, see the original paper\cite{statja25} for details.

A significant disadvantage, however, still exists in the splitting theory: it is impossible to write down the canonical formulation of the theory explicitly (this problem also exists in the original embedding theory).\cite{frtap,davkar}

Nevertheless, one can try to write a path integral by canonical variables in the splitting theory using the method proposed by K. Cahill.\cite{cahill} In the present work we restrict ourselves to the brief outline of the main idea.  

By analogy with the embedding theory\cite{statja18} one can drop out the surface terms and write down the action for the splitting theory (for simplicity we consider only vacuum case):
\begin{align}
S=\frac{1}{2}\int d^{N} y \left(\frac{\dot{z}^A B_{AB} \dot{z}^B}{\sqrt{1+\widehat w_{AB}\dot{z}^A\dot{z}^B}} + \sqrt{1+\widehat w_{AB}\dot{z}^A\dot{z}^B}\, \widehat w^{AB} B_{AB}	\right),
\end{align}
where $\dot{z}^A=\partial z^A/\partial y^0$ and $B_{AB}$, $\widehat w^{AB}$, $\widehat w_{AB}$ are matrices constructed of only spatial derivatives of $z^A$.

Then the generalized momenta have the form:
\begin{align}\label{sp1}
	\pi_A=\frac{\delta S}{\delta \dot{z}^A}=B_{AB}n^B -\frac{1}{2} \widehat w_{AD} n^D \left( n^B B_{BC} n^C - \widehat w^{AB} B_{AB}\right),
\end{align}
where
\begin{align}
	n^A=\frac{\dot{z}^A}{\sqrt{1+\widehat w_{AB}\dot{z}^A\dot{z}^B}}.
\end{align}
In order to obtain a hamiltonian $H(z,\pi)$ we need to express the velocities $\dot z^A$ through the momenta $\pi_A$, but it is impossible in general case because we must solve the multidimensional cubic equation \eqref{sp1} by $n^A$. For the path integral purposes, however, the explicit solution is possibly not nesessary.

First of all, we can write the usual formal expression for the path integral:
\begin{align}
I=<z''|e^{-itH}|z'>=\int Dz D\pi\; \text{exp}\left(i\int d^{N}y(\dot{z}^A\pi_A - H(z,\pi))  \right)	
\end{align}
then perform a formal change of variables\cite{cahill} turning from the integration by momenta $\pi_A$ to $n^A$:
\begin{align}
I=\!\int\!{Dz}{Dn}\left|\frac{\delta \pi_A}{\delta n^B}\right| \text{exp}\left(\!{i}\!\int\!{d}^{N}y\left(\dot{z}^A\pi_A(n) - H(z,\pi(n))\right)\right) \label{int},
\end{align}
where the function $\pi(n)$ is given by \eqref{sp1}, and the hamiltonian $H(z,\	\pi(n))$ can be easily expressed in the explicit form through $z^A$ and $n^A$.

The Jacobian of the above change of variables can be raisen in exponents as usual by introducing ghost variables. As a result, we obtain the path integral of the splitting theory in the explicit form. This integral can be further examined in various approximations, e.g. in the FRW symmetry where $z$ and $n$ become one-dimensional.

\section*{Acknowledgments}
The authors are grateful to the Organizing Committee of IX Friedmann Seminar. The authors would also like to thank K. E. Pavlenko for useful references. The work was partially supported by the Saint Petersburg State University grant 11.38.223.2015.

\end{document}